\documentclass{article}

\usepackage{amsmath} 

\usepackage{arxiv}

\usepackage[utf8]{inputenc} 
\usepackage[T1]{fontenc}    
\usepackage{hyperref}       
\usepackage{url}            
\usepackage{booktabs}       
\usepackage{amsfonts}       
\usepackage{nicefrac}       
\usepackage{lipsum}		
\usepackage{graphicx}
\usepackage[numbers]{natbib}
\usepackage{doi}
\usepackage{algorithm}
\usepackage[algo2e]{algorithm2e} 
\usepackage{algorithmic}
\usepackage{tcolorbox}

\usepackage{setspace} 
\usepackage{subcaption}
\usepackage{tikz}

\usetikzlibrary{shapes,arrows}
\usetikzlibrary{shapes.geometric, arrows, positioning}

\newcommand{\hide}[1]{}

\title{A Misclassification Network-Based Method for Comparative Genomic Analysis}

\author{%
  Wan He \\
  {Northeastern University, Boston, MA, USA}\\
  \texttt{he.wan1@northeastern.edu}
   \And
  Tina Eliassi-Rad \\
  {Northeastern University, Boston, MA, USA}\\
  {Santa Fe Institute, Santa Fe, NM, USA}\\
  \texttt{t.eliassirad@northeastern.edu}
 \And
  Samuel V. Scarpino \\
  {Northeastern University, Boston, MA, USA}\\
  {Santa Fe Institute, Santa Fe, NM, USA}\\
  {Vermont Complex Systems Institute, University of Vermont, Burlington, VT, USA}\\
  \texttt{s.scarpino@northeastern.edu}
}

\date{}

\renewcommand{\headeright}{}
\renewcommand{\undertitle}{}

\hypersetup{
pdftitle={Spatial Analysis of SARS-CoV-2 Sequences via Misclassification Networks},
pdfsubject={q-bio.NC, q-bio.QM},
pdfauthor={Wan He, Tina Eliassi-Rad, Samuel V.~Scarpino},
pdfkeywords={Spatial Analysis, SARS-CoV-2 Sequences, Misclassification Networks},
}

\begin{document}

\maketitle

\begin{abstract}
Classifying genome sequences based on metadata has been an active area of research in comparative genomics for decades with many important applications across the life sciences. Established methods for classifying genomes can be broadly grouped into sequence alignment-based and alignment-free models. The more conventional alignment-based models rely on genome similarity measures calculated based on local sequence alignments or consistent ordering among sequences. However, these alignment-based methods can be quite computationally expensive when dealing with large ensembles of even moderately sized genomes. In contrast, alignment-free approaches measure genome similarity based on summary statistics in an unsupervised setting and are computationally efficient enough to analyze large datasets. However, both alignment-based and alignment-free methods typically assume fixed scoring rubrics that lack the flexibility to assign varying importance to different parts of the sequences based on prior knowledge and also that prediction errors are random with respect to the underlying data generating model. In this study, we integrate artificial intelligence and network science approaches to develop a comparative genomic analysis framework that addresses both of these limitations. Our approach, termed the Genome Misclassification Network Analysis (GMNA), simultaneously leverages misclassified instances, a learned scoring rubric, and label information to classify genomes based on associated metadata and better understand potential drivers of misclassification. We evaluate the utility of the GMNA using Naive Bayes and convolutional neural network models, supplemented by additional experiments with transformer-based models like Enformer \cite{avsec2021effective}, to construct SARS-CoV-2 sampling location classifiers using over 500,000 viral genome sequences and study the resulting network of misclassifications. We demonstrate the global health potential of the GMNA by leveraging the SARS-CoV-2 genome misclassification networks to investigate the role human mobility played in structuring geographic clustering of SARS-CoV-2 and how genomes were misclassified by our model.

\end{abstract}

\section{Introduction}
Artificial Intelligence (AI) based approaches, e.g., models using deep learning architectures, have demonstrated remarkable accuracy in classifying genome sequencing data according to metadata such as geographic sampling location \cite{ounit2015clark,deelder2022geographical}, drug resistance \cite{coll2015rapid,yang2018machine,kuang2022accurate}, taxonomic grouping \cite{murali2018idtaxa,mock2022taxonomic,fiannaca2018deep}, and various other phenotypes of interest \cite{asgari2018micropheno,tampuu2019viraminer,karlicki2022tiara}. Despite the impressive performance of these models, they typically rely on the widely held assumption in machine learning\slash AI that classification accuracy best reflects model performance \cite{brown2024pitfalls}. As a consequence, such AI-based genome comparative methods can likely be enhanced by studying drivers of misclassification. The reasoning is that, while ML\slash AI approaches typically assume errors are driven by random noise inherent in the data, i.e., aleatoric uncertainty, in many biological settings misclassifications are driven by model formulation errors or incomplete representation of the biological system, i.e., epistemic uncertainty. Instead of viewing this epistemic challenge as a weakness, we see an opportunity to derive insights from studying how genomes are misclassified by AI models.

Here, we introduce the Genome Misclassification Network Analysis (GMNA), a genome association analysis framework that extends beyond classification accuracy to extract knowledge from the pairwise relationships among misclassified data. By acknowledging and utilizing the pairwise relationships among the misclassified data, our framework generates additional insights into the underlying processes giving rise to the misclassified genomic data. This framework further allows us to integrate network-science-based approaches into downstream analysis of the drivers of misclassification. We believe the use of the GMNA may provide meaningful insights beyond traditional accuracy-based approaches across the broader field of comparative genomics and for non-genetic data coming from varying application domains.

\paragraph{Comparative genomics:}

Comparative genomics enhances our understanding of the functional and\slash or evolutionary significance of genetic variation by analyzing sequences from different individuals and\slash or species \cite{ulitsky2016evolution,ellegren2008comparative,miller2004comparative,alfoldi2013comparative,frazer2004vista,koonin2000impact}. Among comparative genomics approaches, alignment-based methods such as variants of BLAST \cite{altschul1997gapped,schwartz2003human,kent2002blat} and CLUSTAL \cite{higgins1988clustal,thompson1994clustal,sievers2011fast} utilize dynamic programming algorithms minimizing edit distance \cite{needleman1970general,smith1981identification} to search for aligned sub-sequences. However, such alignment-based methods, which assume collinearity, are limited in capturing higher-order information such as non-linear relationships between subsequences and structural rearrangements within sequences \cite{vinga2003alignment,zielezinski2017alignment}. Moreover, finding the optimal alignment is a combinatorial optimization problem that does not have a continuous solution space to allow for a gradient descent-based optimization scheme \cite{korte2011combinatorial, caramanis2023optimizing}, although prior work suggests alignments based on approximate edit distances should exhibit favourable runtime scaling \cite{bar2004approximating}. Nevertheless, even finding an approximately optimal solution can be computationally infeasible when dealing with even moderate numbers of moderately sized sequences \cite{armstrong2019whole,baichoo2017computational,reddy2024performance}. Alternatively, graph-based alignment methods \cite{kim2019graph,klau2009new} have improved the efficiency and accuracy of aligning whole-genomes, particularly in the highly polymorphic genomic regions. These methods encode genomic sequences as graphs, wherein long, identical subsequences are represented as vertices \cite{angiuoli2011mugsy}. However, this class of methods still faces the challenge of scalability and aligning highly divergent sequences \cite{minkin2020scalable}. Unlike alignment-based methods, alignment-free comparative approaches in genomics typically rely on statistics derived from K-mer words \cite{blaisdell1986measure,reinert2009alignment,wan2010alignment} to capture key features without the need for sequence alignment and are therefore efficient for large-scale genome comparisons. Despite the computational benefits of alignment-free approaches, sequence ordering (including interaction effects) beyond length \textit{k} cannot be modelled by the most K-mer statistics \cite{zielezinski2017alignment}. 

\paragraph{Applications of deep learning to genomics with examples from SARS-CoV-2:}
Deep learning techniques, particularly language models that process sequence data, have significantly impacted genomics-based research \cite{eraslan2019deep, novakovsky2023obtaining}. Here, very high-dimensional models are trained to identify the underlying  patterns in genomic data and have demonstrated impressive performance in downstream tasks such as motif discovery \cite{chaurasia2023human}, promoter region identification \cite{zaheer2020big, ji2021dnabert}, non-coding variant prediction \cite{zhou2015predicting},  protein-ligand binding predictions \cite{chatterjee2023improving}, and gene expression prediction \cite{avsec2021effective}. Among these advancements, transformer-based generative models \cite{vaswani2017attention, devlin2018bert}, such as DNABERT \cite{ji2021dnabert}, BigBird \cite{zaheer2020big}, and Enformer \cite{avsec2021effective} achieved state-of-the-art performance in many applications by integrating long-range interactions through their attention mechanism. However, their training is extremely computationally intensive and constrained by token length. As an example, BERT is quadratically dependent on input sequence length and has a maximum token limit of 512--note that the SARS-CoV-2 genome (which is comparatively small) comprises over 30,000 bases. As a result, SARS-CoV-2 genome sequences cannot be fed into BERT without truncation or serious compression due to their length. While BigBird \cite{zaheer2020big} introduced a block-sparse attention mechanism to reduce complexity to a linear dependency on length, its 4096 token limit still falls short of addressing the full extent of whole genomes. Despite these challenges, AI-driven approaches have been employed to study SARS-CoV-2 in applications such as evolution prediction and lineages identification \cite{wang2023deep,cahuantzi2024unsupervised}
SARS-CoV-2 virus identification \cite{lopez2021classification,han2023predicting},
drug repurposing and prediction
\cite{pham2021deep,jin2021deep,kowalewski2020predicting,beck2020predicting}, diagnosis and treatment \cite{alafif2021machine}.

\paragraph{Genome Misclassification Network Analysis:}
Here, we introduce an alignment-free framework for comparative analysis, which we call the Genome Misclassification Network Analysis (GMNA). The GMNA measures associations between ensembles of genome sequences based on the empirical likelihood of their misclassification. The discriminative filters used to compare the genome ensembles are learned by an arbitrary AI model using genome feature information. Briefly, the GMNA starts with a trained classifier that takes as input a genome sequence and predicts metadata of interest. We then construct a weighted misclassification network using the misclassified instances, where nodes represent ensembles of genome sequences from different metadata classes. Edges indicate associations between pairs of genome sequence ensembles, with weights representing "indistinguishability" scores derived from empirical misclassification likelihoods between ensembles from distinct metadata classes. By selecting specific features of interest, pairwise associations among genome sequences with different realizations of these feature variables can be explored. GMNA offers a computationally efficient comparative genomics tool for studying associations between large ensembles of genome sequences while leveraging valuable label information. 

To address limitations of the alignment-based comparative genomics approaches, instead of attempting to explicitly design a quantity to measure the association among genome sequences, we propose a framework that could use any arbitrary AI model to learn discriminative features and quantify the association between ensembles of genome sequences. Under this AI-based misclassification network framework, we rely on the judgement of the machine to determine how close a pair of genome ensembles are by computing an \emph{indistinguishability} score, based on the empirical misclassification likelihood. The GMNA framework hypothesizes that--given a decent classifier--misclassification happens when the predicted class and the true class of the sequences shared sufficient distinctive features and therefore are \emph{indistinguishable} enough to deceive the AI to misidentify the input sequences.

Without manual feature engineering or explicitly specifying a feature selection scheme for the genome classification task, the AI model automatically learns the features that optimize performance. Consequently, using the AI framework also has the added benefit of incorporating target output, i.e., the class labels, in the feature mining process. Moreover, this framework offers flexibility in the choice of the appropriate AI model. The classification model could be tailored to suit the specificity of the learning problem. For instance, a binary genome sequence classification problem can be accurately solved with nearly 100$\%$ accuracy using a convolutional neural network. However, under computation resource constraints especially for more complex tasks, such as fine-grained sequence classification on longer sequences, a Naive Bayes model may be a more efficient choice. We emphasize that the purpose of this study is not to advance a specific model's classification performance on whole genome sequence,  but to offer an alternative and efficient approach for genome ensemble comparative analysis based on the often overlooked misclassified instances. 

To demonstrate the utility of the GNMA, we utilized 551,230 labeled whole genome SARS-CoV-2 sequences, where each label denotes the specific region in which the sequence was collected. We used CNN and Naive Bayes as the backbone methods of our framework. With spatial sampling location predictions from these trained classifiers, we compared the learned misclassification network with a flights network from Official Airline Guide (OAG) to study how travel impacts genome variation.  Finally, we discuss how our GMNA may be applied to datasets labeled with multiple features.

\subsection*{Contributions}

The key contributions of this work are as follows:

\begin{itemize}
    \item We introduce a novel alignment-free framework, termed the Genome Misclassification Network Analysis (GMNA), that uses empirical misclassification likelihoods to measure associations between ensembles of genome sequences.
    
    \item We identify the pairwise association between the target outcome and the predicted outcome, which is then utilized to design a data-driven framework that can incorporate any arbitrary AI model for use in comparative genome analysis.

    \item We propose the concept of \textit{indistinguishability} as a metric to quantify the association between pairs of genome groups, which captures the genetic diversity and complexity of genome ensembles.

    \item Our framework is adaptable and can incorporate any AI model for comparative genome analysis, making it applicable across various genomic datasets and features.
    
    \item Using more than 500,000 SARS-CoV-2 genomes, we demonstrate that GMNA can uncover geographic associations between genome sequences with limited computational resources.
    
    \item We construct a misclassification-based network and compare it with the Official Airline Guide (OAG) flights network, highlighting how human activities, such as global transportation, impact genome variation and evolution.
    
    \item We introduce the leave-one-class-out (LOCO) model to balance the trade-off between generating sufficient misclassified data for association inference and maintaining model credibility with high prediction accuracy.
    
    \item Our framework provides a computationally efficient tool for large-scale comparative genomics, offering insights into phylogenetic structure and evolutionary patterns.
\end{itemize}

\section{Genomic Misclassification Network Analysis (GMNA)}
\label{sec:GMNA}

\begin{figure}[htbp]
\centering
\includegraphics[width=0.8\textwidth]{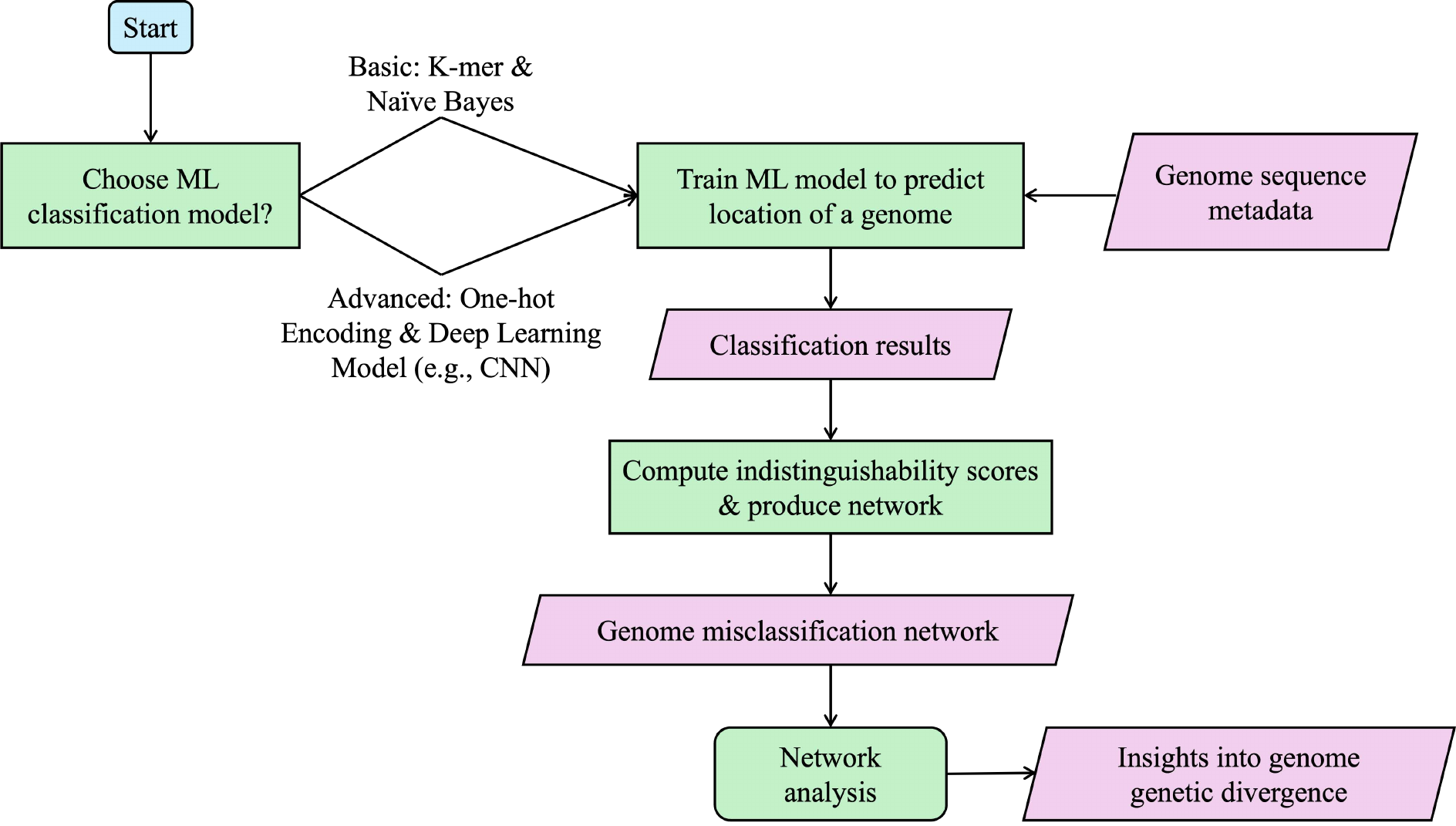}
\caption{Flowchart of the Genome Misclassification Network Analysis Framework.}
\end{figure}

We designed an integrative procedure that combines AI and methods from network science to perform comparative genomic analysis. We first solved a genome sequence classification problem, mapping genome sequences to metadata. The misclassified instances outputted from the classifying machine are then utilized to compute \emph{indistinguishability} scores that measure pairwise associations among genome groups for correlation-based analysis. We interpret the edges in this genomic misclassification network as the association between ensembles of genome sequences under the assumption that genomes with more shared discriminative latent features--and thus fewer divergent features--are more likely to confuse the machine and mislead it into making mistakes. In other words, the more \emph{indistinguishable} a pair of genome classes appears to the trained AI, the more similar the two classes are. For example, if we are trying to predict spatial sampling location, we set the target class labels to be regions where genome sequences were collected. A misclassification network can then be constructed from the classification results, where nodes are the genome collection regions, and an edge between region A and region B is established when a genome from region A is incorrectly classified as being from region B. 

More generically, each node in a misclassification network represents data of a specific class where each class is one of the instances in the target label set $\mathcal{Y}$. The target set $\mathcal{Y}$ for the classifier to predict is chosen to be the attribute of which the association among the instances is the subject of research interest. Associations between any pair of data ensembles labeled by class instances $(c_{i}, c_{j})$ of the attribute of research interest $\mathcal{Y}$ are measured by how indistinguishable these data points are to the classifying machine. The more indistinguishable the data labeled by these two classes appears to the AI, the more similar we can assume the classes to be. Formally, the weight $w_{ij}$ of edge $e_{ij}$ in the misclassification network, interpreted as the association between class $c_{i}$ and $c_{j}$ $\in \mathcal{Y}$, is given by the \emph{indistinguishability} score between data ensembles $\mathcal{X}_{i}$ and $\mathcal{X}_{j}$, where $\mathcal{X}_i=\{x \in \mathcal{X}: y(x)=c_i\}$, computed as the symmetrized empirical probability of misclassification, as shown in Equation \ref{eq:indistinguishability}. When applied to genomic data we call our framework, the Genome Misclassification Network Analysis (GMNA). Despite focusing on genomic data for the remainder of the paper, we stress that the underlying methodology is quite general and can be readily extended to non-genomic data. 

The selection of suitable AI models within the GMNA framework depends on the complexity of the classification problem at hand and resolution of the metadata. For example, a challenging aspect of classifying SARS-CoV-2 whole genome sequences is the high computational cost associated with their length. SARS-CoV-2 genomes are around 35,000 bases and are composed of a very small vocabulary of 16 base letters, which includes degenerate bases. The long length relative to vocabulary size makes it critical for the classifier to account for combinations, permutations, local context, and long-range interactions among subsequences, as genome sequences exhibit a highly context-sensitive structure. To capture the essential subsequence patterns within the genomic data, we adopted K-mer preprocessing coupled with Naive Bayes as a baseline approach for GMNA. The K-mer preprocessing allowed the expansion of the vocabulary used in genome sequence analysis and thus increased the diversity of features considered while incorporating local context. This choice was also motivated by the method's computational efficiency, which remains independent of the sequence length or the number of sequences. Importantly, this independence allows us to efficiently measure associations between large ensembles of lengthy genome sequences, making our approach robust and scalable. To enhance the classifier's performance, we employed a convolutional neural network (CNN) architecture. The CNN reduces sequence length while learning filters to extract label-specific, discriminative features from the genomes, enabling efficient and automatic feature extraction.

The resulting misclassification network could then be further analyzed using any network analysis techniques to study its network properties of interest. For instance, it could be fed to any community detection algorithms \cite{girvan2002community, 6729613, halkidi2001clustering,leskovec2010empirical,fortunato2010community} to evaluate how data with different attributes of interests are associated. Alternatively, nodes, i.e. genome ensembles, could be characterized based on any centrality measure such as betweenness \cite{newman2005measure} or page-rank \cite{page1999pagerank}.

\subsection{Misclassification Network Construction}
\subsubsection{Problem definition}
To obtain the pair-wise associations based on \textit{indistinguishability} scores from the misclassification results, we consider the classification problem of mapping a set of N genome sequences $\mathcal{X}=\{x_1,..., x_N\}$ to their location labels $\mathcal{Y}=\{loc(x_1),..., loc(x_N)\}=\{y_1,..., y_N\}=\{r_1,..., r_K\}$. Here, $K$ represents the cardinality of the label set $\mathcal{Y}$. We define $\mathcal{X}_{r_i}$ as genomes originating from region $r_i$, where $r_i \in \mathcal{Y}$ and $\mathcal{X}_{r_i}=\{x_m|\  y_m=r_i,\ m\in \{1,..., N\}\}$. Denote $E_f(r_i,r_j)$ as the event of a classifier \emph{f} incorrectly classifying a genome sequence from region \emph{i} to region \emph{j}. Denote $\mathcal{C}_f(\mathcal{D}_s,r_i,r_j)$ as the count of occurrences of event $E_f(r_i,r_j)$ where $r_i,r_j \in \mathcal{Y}$  when input data $\mathcal{D}_s$ is used for predictions. Formally, $\mathcal{C}_f(\mathcal{D}_s,r_i,r_j)=\sum_{x\in\mathcal{X}_{r_i}\subseteq\mathcal{D}_s} I(f(x)=r_j|y(x)=r_i)$. The empirical probability of sequences from region $r_i$ being misclassified as from region $r_j$ by classifier \emph{f} is calculated as  $P_f(\mathcal{X}_{r_i},r_i,r_j) = \frac{C_f(\mathcal{X}_{r_i},r_i,r_j)}{|\mathcal{X}_{r_i}|}$. The objective function for classifier training is to minimize the total misclassification cost:
\begin{equation}
\sum_{x\in\mathcal{X}} cost(f(x),y(x)).
\end{equation}

We summarized our notation in Table~\ref{tab:GMNA_notation}.
\begin{table}[htbp]
\centering
\caption{Notation Table}
\begin{tabular}{|c|p{10cm}|}
\hline
\textbf{Symbol} & \textbf{Description} \\
\hline
$N$ & Total number of genome sequences \\
$K$ & Cardinality of the label set $\mathcal{Y}$ \\
\hline
$\mathcal{X}$ & Set of $N$ genome sequences, $\mathcal{X}=\{x_1, \ldots, x_N\}$ \\
$\mathcal{Y}$ & Set of location labels, $\mathcal{Y}=\{loc(x_1), \ldots, loc(x_N)\} = \{y_1, \ldots, y_N\} = \{r_1, \ldots, r_K\}$ \\
\hline
$\mathcal{X}_{r_i}$ & Set of genomes originating from region $r_i$, where $r_i \in \mathcal{Y}$ \\
& $\mathcal{X}_{r_i}=\{x_m \mid y_m=r_i, m \in \{1, \ldots, N\}\}$ \\
\hline
$E_f(r_i,r_j)$ & Event of classifier $f$ incorrectly classifying a genome sequence from region $i$ to region $j$ \\
\hline
$I(f(m)=r_j\mid y(m)=r_i)$ & Indicator function evaluating to 1 if classifier $f$ predicts $r_j$ given that the true label is $r_i$ for genome sequence $m$, otherwise 0 \\
\hline
$\mathcal{C}_f(\mathcal{D}_s,r_i,r_j)$ & Count of occurrences of event $E_f(r_i,r_j)$ when input data $\mathcal{D}_s$ is used for predictions \\
& $\mathcal{C}_f(\mathcal{D}_s,r_i,r_j)=\sum_{m\in\mathcal{X}_{r_i}\subseteq\mathcal{D}_s} I(f(m)=r_j\mid y(m)=r_i)$ \\
\hline
$P_f(\mathcal{X}_{r_i},r_i,r_j)$ & Empirical probability of sequences from region $r_i$ being misclassified as from region $r_j$ by classifier $f$ \\
& $P_f(\mathcal{X}_{r_i},r_i,r_j) = \frac{C_f(\mathcal{X}_{r_i},r_i,r_j)}{|\mathcal{X}_{r_i}|}$ \\
\hline
\end{tabular}
\label{tab:GMNA_notation}
\end{table}

Our proposed Misclassification Network Analysis (MNA) framework can be applied under various setups. Below, we discuss a few options:

\paragraph{Binary Misclassification Network Analysis (MNA)} \label{paragraph:BinaryMNA}
Under the binary MNA setting, pairwise genome ensemble similarities are obtained by training a binary classifier for every pair of genome ensembles. The binary-classification setting computes the pairwise genome ensemble association between $\mathcal{X}_{r_i}$ and $\mathcal{X}_{r_j}$ using a training dataset containing only genome sequences from the two regions $r_i$ and $r_j$. The association between any two ensembles is measured by the inverse of the classification accuracy of the binary classifier, as shown in Equation \ref{eq:indistinguishability}.

\paragraph{Multiclass MNA}\label{paragraph:MCC_MNA}
In the multi-class MNA setting, pairwise genome ensemble similarities are obtained by training a multiclass classifier (MCC) and aggregating the misclassified instances. Thus computation of inter-class associations among all pairs of data ensembles requires training only one classifier. For GMNA, the \emph{indistinguishability} scores of all pairs of genome ensembles $\mathcal{X}_{r_i}, \mathcal{X}_{r_j}$ become available once the MCC model completes training and makes predictions on the test set. 

Under both the binary and multiclass settings, the number of nodes for misclassification network analysis (MNA) is given by the order of the target variable set $\mathcal{Y}$, specifically $\mathcal{Y}_{train} \cap \mathcal{Y}_{test}$, that is the number of classes for prediction. For spatial analysis on the genome sequences, the target variable is the location information of the genomes; thus, each node represents genome sequences ensembled by the region labels.  The \emph{indistinguishability} score between genome ensemble $\mathcal{X}_{r_i}$ and $\mathcal{X}_{r_j}$ from classifier $f$ is given by:
\begin{equation}
\label{eq:indistinguishability}
indistinguishability(\mathcal{X}_{r_i},\mathcal{X}_{r_j})=P_{f}(\mathcal{X}_{r_i},r_i,r_j)+P_{f}(\mathcal{X}_{r_j},r_j,r_i).
\end{equation}

\paragraph{Leave-one-class-out (LOCO) MNA}
\label{sec:LOCO}

However, both the binary and MCC misclassification network analysis (MNA) schemes face a trade-off between prediction accuracy and having enough misclassified instances to establish meaningful correlations for inference. A high-performance classifier accurately categorizes genome sequences has the issue of not producing sufficient misidentified data for class-wise association inferences. To address this dilemma, we propose the leave-one-class-out (LOCO) misclassification generation scheme. Although in this work, our goal is not to eliminate all the incorrectly classified cases, since the purpose of this study is not to push classification accuracy, but rather to study the temporal and spatial associations among SARS-CoV-2 genome sequences based on the incorrectly classified cases. However, a poorly performing AI that frequently misidentifies the genome origins lacks credibility for further results analysis. 

To resolve the dilemma between consistently generating more incorrectly classified data points for inference and maintaining satisfactory prediction accuracy, we introduced a leave-one-class-out (LOCO) model. This approach iteratively leaves out one class of genome sequences during the training process, and the model is subsequently trained to classify the excluded sequences, and is trained to later classify the genome sequences that were left out during training. The LOCO classifier appoints a "centroid" region and trains a classifier on a data subset that excludes genome sequences from this selected centroid region. This process iterates through the entire set of regions, with each region taking a turn as the centroid.  After training, the leave-one-class-out classifier is applied to the genome sequences from the currently excluded centroid region. One major benefit of this setting is the guaranteed abundance of misclassified instances and thus greater statistical significance in general, since each genome from the centroid is inevitably misclassified into one of the remaining classes, as the training subset does not include the true target region (the centroid itself). This setting produces a star misclassification network, with the centroid as the hub and the degree of a node corresponding to the frequency of genome sequences from the centroid region getting misclassified as originating from the region corresponding to that node. Finally, the genome ensemble association between the centroid and the remaining regions could be computed as,

\begin{equation}
indistinguishability(\mathcal{X}_{centroid},\mathcal{X}_{r_j})=P_{LOCO}(\mathcal{X}_{centroid},r_{centroid},r_j)
\end{equation}

For example, if the US is the appointed left-out "centroid" region, the leave-one-class-out classifier will be trained on the entire preprocessed dataset excluding the genome sequences from the US. During prediction, the test set will consist only of the US sequences, and the leave-one-class-out classifier will be used to map the US genomes to a non-US country. As a result, all the predictions made are incorrect and are used solely for correlational inference. Under this setting, which produces a consistent amount of incorrectly classified samples, improving the classification accuracy no longer limits the amount of data needed for the misclassification analysis.

\begin{tcolorbox}[colframe=black, colback=white, sharp corners, boxrule=1pt, boxsep=0pt]

\begin{algorithm}[H]
\caption{LOCO Algorithm}
\begin{algorithmic}
\caption{LOCO Algorithm: This algorithm addresses the dilemma between the availability of misclassified instances for inference and maintaining prediction accuracy by purposefully generating misclassified instances.}

\STATE \textbf{Input:}
\STATE $\mathcal{X}_{\text{centroid}} \in \{0,1\}^{B \times L_{max} \times N_{\text{centroid}}}$: One-hot encoded genomes from the centroid region, where $B$ is the number of nucleotide bases, $L_{max}$ is the length of the longest genome, and $N_{\text{centroid}}$ is the number of genomes in the centroid region

\STATE \textbf{Output:}
\STATE \begin{enumerate}
    \item $f_{\text{LOCO}}^{\text{centroid}}$: Trained classifier for the centroid region
    \item $\text{indistinguishability}(r_{\text{centroid}}, r')$: Association measure between centroid region and region $r'$
\end{enumerate}

\FORALL{region $r$}
    \STATE \begin{enumerate}
        \item centroid region $r_{centroid}$ := \emph{r}
        \item $\mathcal{X}_{train} := \mathcal{X}_{train}-\mathcal{X}_{centroid}$
        \item Train classifier $f_{LOCO}^{centroid}$ that minimise $\sum_{x\in\mathcal{X}_{train}} cost(f(x),y(x)) p(x)$
    \end{enumerate}

    \FORALL{region $r' \neq r$}
        \STATE Compute "association" between genome ensembles from centroid region and region $r'$ as:
        \[
        \text{indistinguishability}(r_{\text{centroid}}, r') = \frac{\sum_{x_i \in \mathcal{X}_{\text{centroid}}} I(f_{\text{LOCO}}(x_i)=r_j)}{|\mathcal{X}_{\text{centroid}}|}
        \]
    \ENDFOR
\ENDFOR
\end{algorithmic}
\end{algorithm}
\end{tcolorbox}

\paragraph{Soft misclassification}\label{sec: soft classification}
As an alternative to using the final prediction results to compute the \emph{indistinguishability} score, we could instead take a step back by considering the entire distribution from the last activation function, which computes a probability for each possible realization in the output set $\mathcal{Y}$. In the soft classification setting, we could have the classifier output the entire multinomial distribution, consider the Softmax activation function as an example, $\big\{softmax_f(y|x_i)\big\}_{y\in\mathcal{Y}}$ as opposed to just the label with the maximum score $y^*:=max_y\big\{P_f(y|x_i)\big\}_{y\in\mathcal{Y}}$. Therefore the misclassification distribution for computing \emph{indistinguishability} could also be computed as:

\begin{equation}
    P_f(\mathcal{X}_{r_i},r_i,r_j) =\frac{\sum_{x\in\mathcal{X}_{r_i}} softmax_f(r_j|x)}{|\mathcal{X}_{r_i}|}
\end{equation}

\paragraph{Aggregated Indistinguishability}

The $\emph{aggregated indistinguishability}$ for a region $r_i$ is defined as the sum of the pairwise indistinguishability between $r_i$ and all other regions $r_j \neq r_i$:
\[\text{Aggregated Indistinguishability}(r_i) = \sum_{j \neq i} \text{Indistinguishability}(\mathcal{X}_{r_i}, \mathcal{X}_{r_j}),
\]
Substituting the definition for pairwise indistinguishability, the aggregated indistinguishability becomes:
\begin{equation}
\text{Aggregated Indistinguishability}(r_i) = \sum_{j \neq i} \left[ P_f(\mathcal{X}_{r_i}, r_i, r_j) + P_f(\mathcal{X}_{r_j}, r_j, r_i) \right].
\end{equation}

\paragraph{Connection to TPV and PPV}

The True Positive Value (\( \text{TPV} \)) and Positive Predictive Value (\( \text{PPV} \)) for region \( r_i \) could be written as:
\begin{align*}
\text{TPV}_{r_i} &= 1 - \sum_{j \neq i} P_f(\mathcal{X}_{r_i}, r_i, r_j), \\
\text{PPV}_{r_i} &= 1 - \sum_{j \neq i} P_f(\mathcal{X}_{r_j}, r_j, r_i).
\end{align*}

Thus, the aggregated indistinguishability for region \( r_i \) is equivalently given by:
\begin{equation}
\text{Aggregated Indistinguishability}(r_i) = 2 - (\text{TPV}_{r_i} + \text{PPV}_{r_i}).
\end{equation}

\section{Experiments and Results}

\subsection{Data and preprocessing}\label{subsec: Data and preprocessing}

\paragraph{Data}
For our study, we used a dataset of 551,230 SARS-CoV-2 genome sequences with a maximum length 34,692 from 198 regions. The genome data recorded the SARS-CoV-2 genome sequences along with the location where and the date when they were discovered. Data were obtained from the GISAID database and used as per the GISAID Terms of Use. For a list of genomes, acknowledgement of Originating and\slash or Submitting Laboratory please see Supplemental Table \ref{tab:sup-gisaid}. 

\paragraph{Preprocessing}
This dataset presents two major challenges that affect its analysis: data imbalance and computation complexity brought by the length of the genome sequences. Specifically, the data availability is significantly more ample in the UK, the US and Italy than the rest of the world and it also include minority classes with very few datapoints to learn their patterns. Secondly, the average whole genome sequence length is over 30,000 and the lack of pre-trained base-model for whole genome sequence data makes training of the deep learning neural network computationally expensive and prone to overfit. 

To address these challenges, for our experiments, an equal number of genomes are sampled from countries whose data availability passed a certain threshold, resulting in a dataset with equal regional representation. The smaller this threshold, the larger the output set for the MCC problem. The sampling threshold and the sample size from each region, determines the data size and hence computational complexity of solving the classification problem. Hence, the exact architecture used for the classification problem varies with the experiment setting, which in turn determines the computation complexity of the problem. 

Further, to investigate the evolutionary timeline of the SARS-CoV-2 genomes, such as identifying the point of evolutionary divergence, our experiment involved a temporal analysis that considered genome association variation at varying timepoints to understand how the virus has evolved over time, enabling us to trace back to the periods of significant genetic shifts and to assess the pace of the evolution.

\paragraph{Baseline: kmers for Naive Bayes classification}
We chose Kmer pre-processing coupled with Naive Bayes as a baseline approach for GMNA since its computation cost is independent of the length of the sequences. For kmer-based Naïve Bayes modelling, the genome sequences were first transformed into sequences of Kmer words of which the letters are the nucleotides. The K-mer words pre-processing could be considered as using a sliding window of size k to scan through the genome sequences of size L to obtain a sequence of size L-k+1 composed of words of length k. The Naive Bayes model assumes Bayes' theorem and conditional feature independence, in this case the Kmer genome subsequence frequencies are assumed to be conditionally independent given the region class labels.

\paragraph{Padding and one-hot encoding for DNN}
We determined the length of the longest sequence in the genome set and padded all sequences to this maximum length. The padded sequences are then one-hot encoded to obtain a numerical representation of the genome sequences that could be fed to a neural network. After this transformation, the data is represented by a tensor of shape \emph{max length} $\times$ \emph{(number of distinctive bases +1)} $\times$ \emph{number of genome sequences}.

\subsection{Evaluation}

The performance of the framework could be assessed from four aspects:

\paragraph{Qualitative comparison with the ground truth}
The alignment between the implications from the misclassification network and expected outcome based on reality or prior knowledge could also be used to validate the efficacy of the MNA framework. For example, in this work, we are expecting the genomes to exhibit spatial dependencies based on the assumption that traveling and social contacts have a significant impact on the \emph{indistinguishability} between genome ensembles. By considering the process of disease spreading and hence genome replications as a random walk, we hypothesize that the degrees of separation between genomes is positively correlated to the probability of mutations during RNA replications. To define the degrees of separation in this context, if a person A contracted the disease from B and similarly B from C, then the genome from A is one degree from the genome from B and 2 degrees away from the genome from the person C. Meanwhile, we assume the degrees of separation among hosts and hence genomes are inversely related to their geographical proximity. Thereby, we expect genomes found in neighboring regions to exhibit stronger associations and hence greater \emph{indistinguishability}.

\paragraph{Classification performance}
The classification performance, e.g., prediction accuracy, of the trained machine could be considered as an indicator of the intelligence of the machine. The more intelligent the machine, the more credible the suggested association is between the incorrectly predicted class and the true class.

\paragraph{Robustness to the choice of models and classifier accuracy}
The robustness of the results could be assessed by comparing the misclassified instances produced by different machine learning models and their sensitivity to the classifier accuracy. The inferences obtained from the misclassification network should be resistant to changes in the ML model used for classification. On the other hand, if the misclassification network produced by a weak classifier is consistent with that produced by a strong classifier, the results are considered to be more stable.

\paragraph{Comparison with its configuration network}
The significance of the results could be assessed by comparing the obtained misclassification network with is configuration network that preserves the same weighted degree sequence, but permuting the connections within the network. This means that the regions they got misclassified as are shuffled while keeping the misclassification probability of each region constant. The configuration model could be used as a null model to analyze whether the identified patterns in the true misclassification network could also be found in its configuration network where all network structures except for the connections between nodes are randomized and perturbed. This comparison helps to determine whether the observed patterns are merely a result of the indistinguishably of each genome ensemble. 

\subsection{Results}

\subsubsection{Spatial Analysis: SARS-CoV-2 genomes exhibit strong spatial dependencies}
\begin{figure}[htbp]
\centering
\begin{subfigure}[b]{1\textwidth}
   \includegraphics[width=1\linewidth]{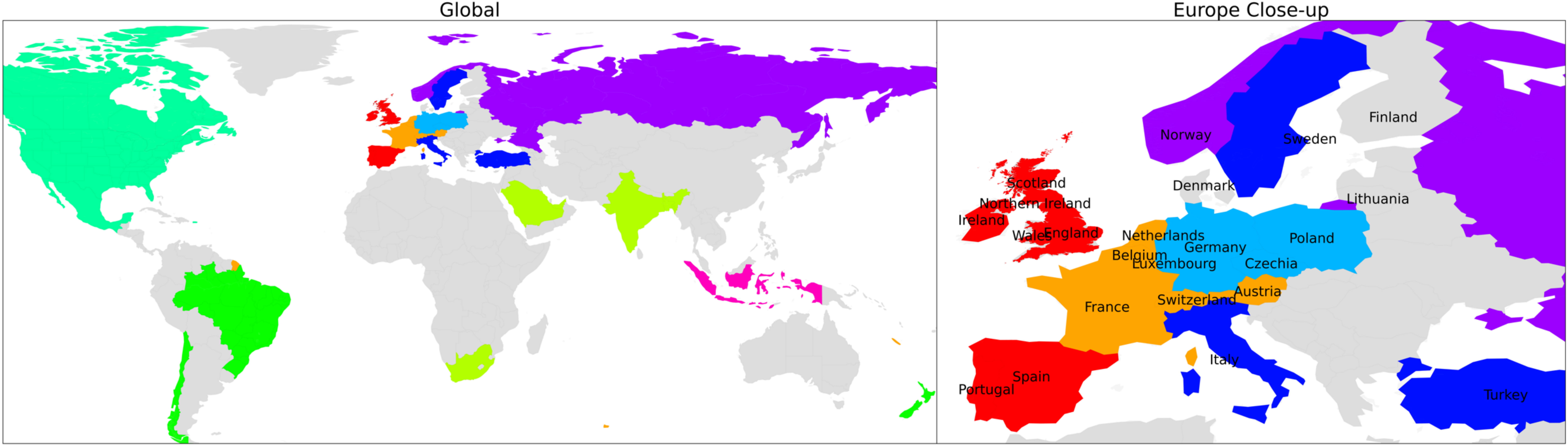}
   \caption{Genome Misclassification Results Under the Multiclass Setting: Regions are colored based on Louvain community detection applied to the misclassification network, where the same color represents regions within the same community. A Naive Bayes multiclass model was used with a sample size and threshold of 300. Singular communities (one region) are annotated but left uncolored to distinguish them from larger groupings.
   }
   \label{subfig:MCCNBtrue}
\end{subfigure}

\begin{subfigure}[b]{1\textwidth}
   \includegraphics[width=1\linewidth]{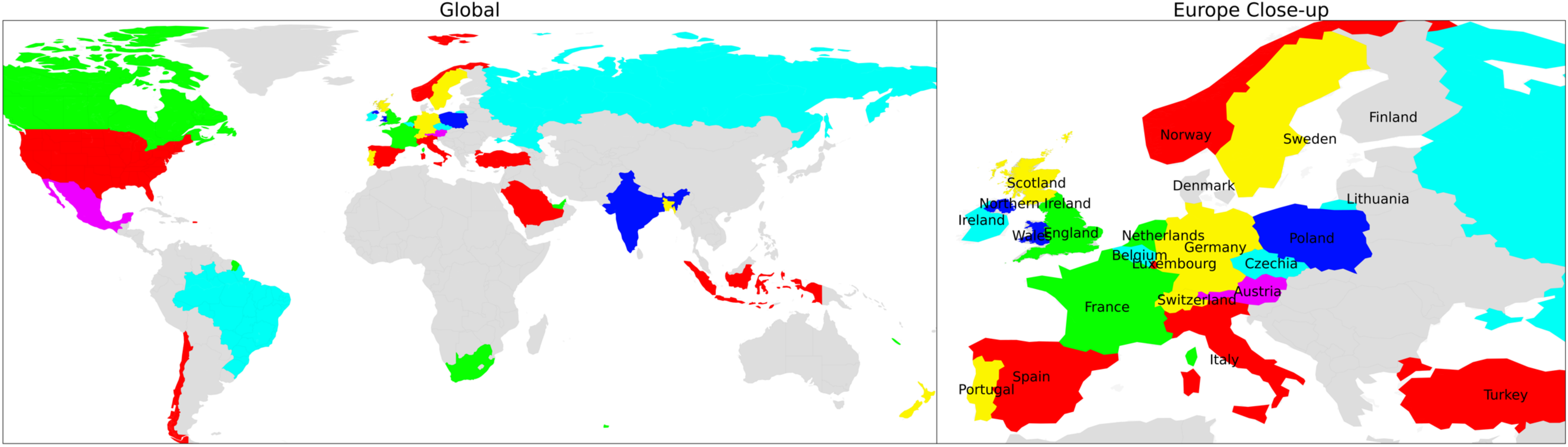}
   \caption{The configuration network of the The community partition of the configuration model is shown, where the degree (indistinguishability score) of each region is preserved, but the connections are randomized to assess the significance of the spatial dependencies observed in (a).}
   \label{subfig:MCCNBconfig}
\end{subfigure}
\caption{MCC Misclassification vs. Configuration Clustered by the Louvain Community Detection Algorithm: Our results show the SARS-CoV-2 genomes exhibit strong spatial dependencies. Genomes from neighboring regions are highly indistinguishable, and the significance of the findings is supported by the configuration model test.}
\label{fig:MCCNB}
\end{figure}

Our results demonstrate the strong spatial dependencies in SARS-CoV-2 genomes, as shown by the community detection results on the misclassification network. Firstly, we demonstrate the applicability of our proposed framework MNA under the multiclass setting~\ref{paragraph:MCC_MNA}. We sampled 300 sequences from each region to ensure broad representation and inclusion of most regions in the analysis. To build the misclassification network, classifiers are trained to predict the origin of genome sequences as outlined in Sec.~\ref{sec:GMNA}. The pairwise relations between the true and predicted classes are aggregated to construct the misclassification network, which is then partitioned by the Louvain community detection algorithm. The resulting communities, visualized in Fig. \ref{subfig:MCCNBtrue}, reveal that regions with close geographic proximity around the globe are often partitioned into the same community by the community detection algorithm, which means that the genome sequences from neigboring regions are more indistinguishable from each other. These results support our hypothesis that misclassified instances capture meaningful correlations and could be utilized for inference. Genomes from geographically close regions are more indistinguishable, supporting the assumption that travel and social interactions contribute to genome similarity.

\begin{figure}[htbp]
\centering
   \includegraphics[width=\linewidth]{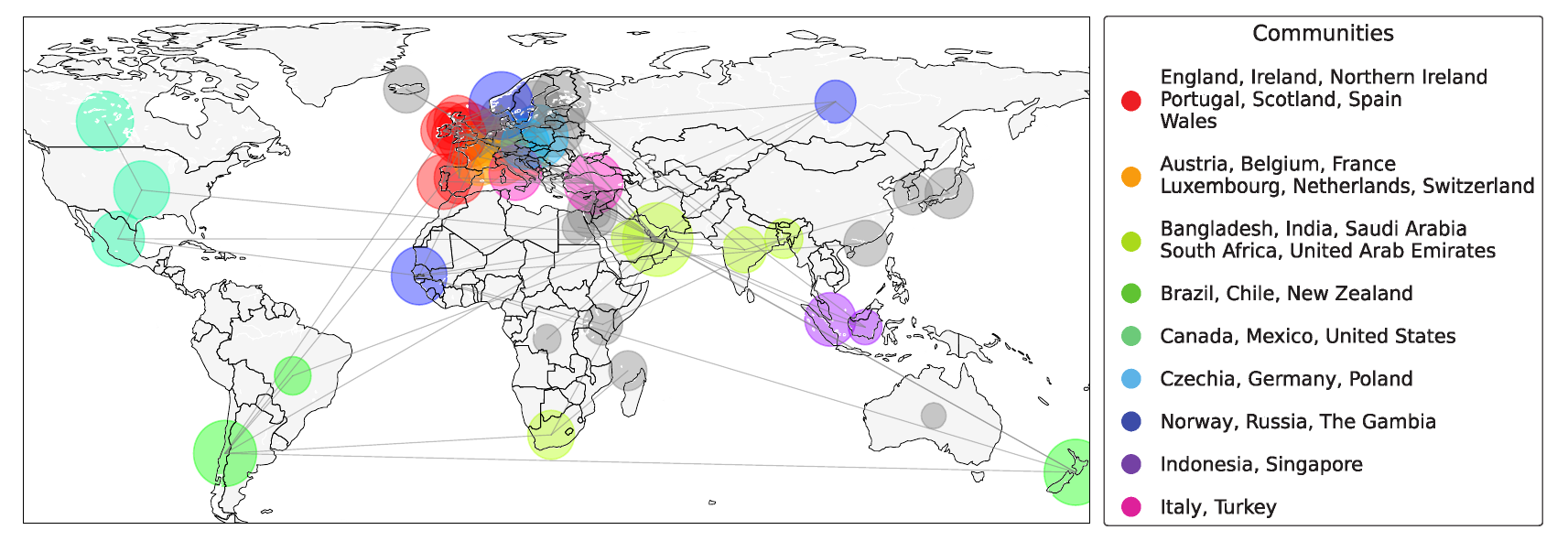}
\caption{Network Visualization of Misclassification Results in Fig. \ref{subfig:MCCNBtrue}: In this genome misclassification network, each node represents a subset of genome sequences from a specific region, with edge weights representing the indistinguishability score, computed as the symmetrised empirical likelihood of misclassification between genomes from the two regions. The partition and coloring of the regions are based on community detection clustering results, with singular communities consisting of only one region colored in grey. This plot shows the top 20\% edges with the greatest weights.}
\label{fig:MCCNB_network}
\end{figure}

In the true misclassification network in  Fig.~\ref{subfig:MCCNBtrue}, Ireland and the UK's four constituent countries along with Spain and Portugal form a distinct community. Other neighboring regions grouped together include: Canada, the United States and Mexico in North America; Brazil and Chile in South America; Germany, Czech Republic, Poland, Luxembourg in Central Europe; Belgium, Denmark, France, the Netherlands, and Austria in Northwestern Europe; Saudi Arabia, United Arab Emirates, India, Bangladesh, and South Africa in the Middle East, South Asia, and Southern Africa; and Indonesia and Singapore In Southeast Asia. Additionally, we provide a network visualization of the GMNA results in Fig.\ref{fig:MCCNB_network}.

The configuration model provides a null hypothesis by preserving regional genome indistinguishability scores but permuting the connections. Comparing the observed misclassification network with the configuration model (Fig. \ref{subfig:MCCNBconfig}) allows us to assess whether the observed spatial dependency patterns are significant or merely artifacts.

Secondly, we demonstrate MNA under the binary setting (see ~\ref{paragraph:BinaryMNA}), using selected centroid countries from different continents. As described in ~\ref{paragraph:BinaryMNA}, the resulting misclassification network from the binary classification setting is a star network, with edges connecting the centroid country to the rest of the world. The edge weights represent the indistinguishability between the centroid and the other regions, computed by training a binary classifier on genome sequences from the centroid region and each of the other regions, and then calculating an indistinguishability score based on the misclassification results. 

\begin{figure}[htbp]
    \centering
    \begin{subfigure}{1\textwidth}
        \centering
        \includegraphics[width=0.8\linewidth, height=1.8in]{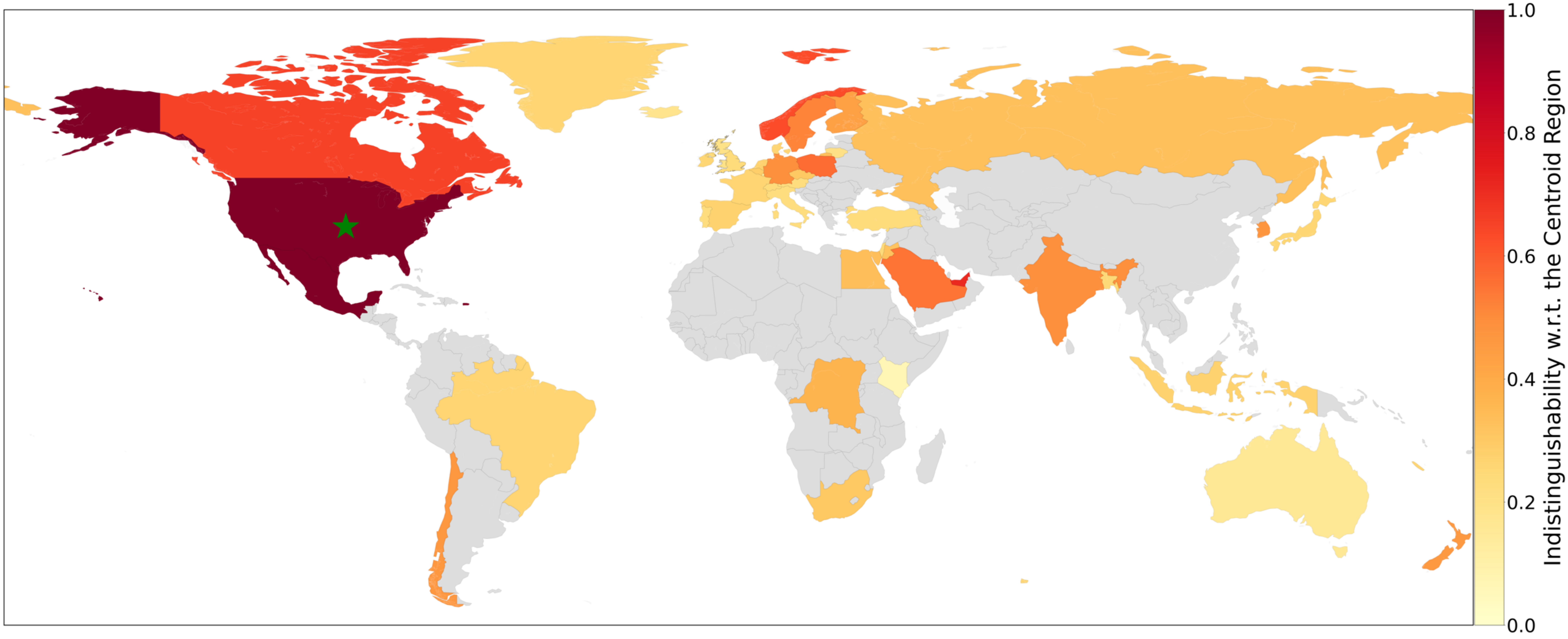}
        \caption{Centroid Region= United States}
        \label{fig:plot1}
    \end{subfigure}
    \\
    \begin{subfigure}{1\textwidth}
        \centering
        \includegraphics[width=0.8\linewidth, height=1.8in]
        {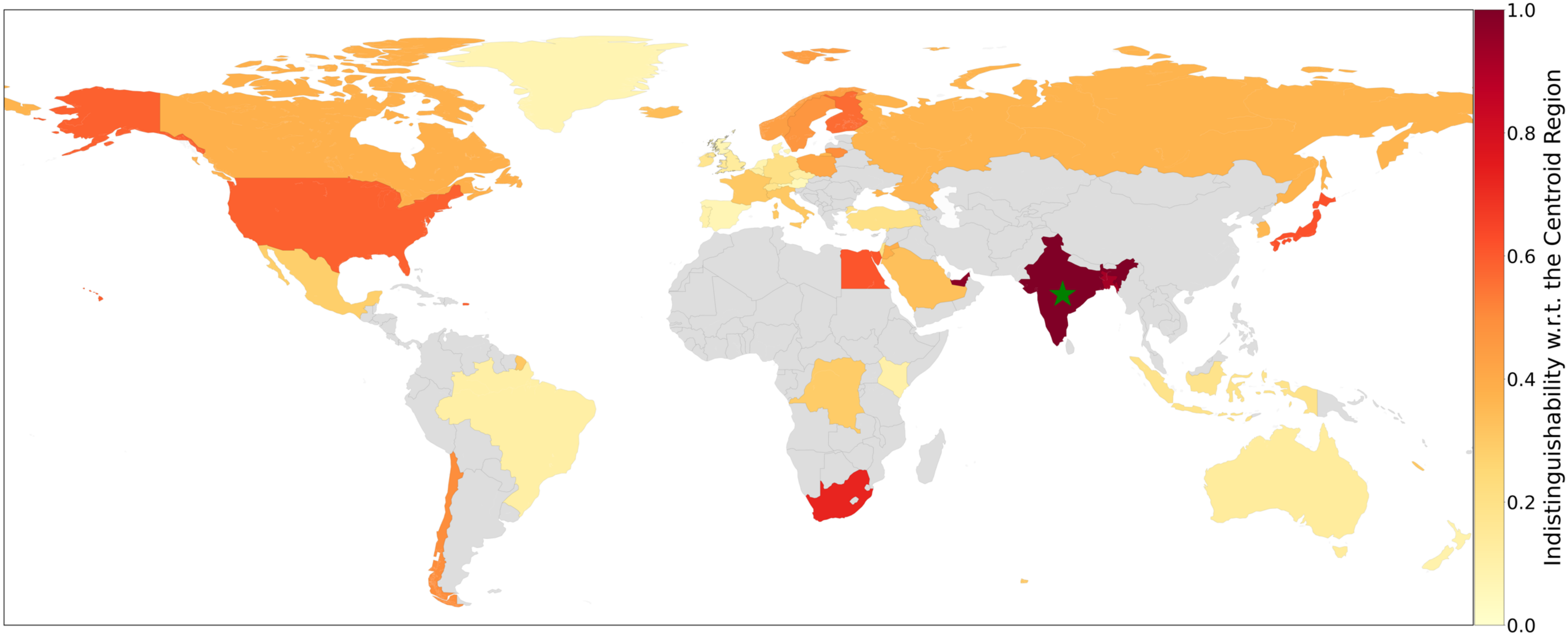}
        \caption{Centroid Region= India}
        \label{fig:plot2}
    \end{subfigure}
    \\
    \begin{subfigure}{1\textwidth}
        \centering
        \includegraphics[width=0.8\linewidth, height=1.85in]{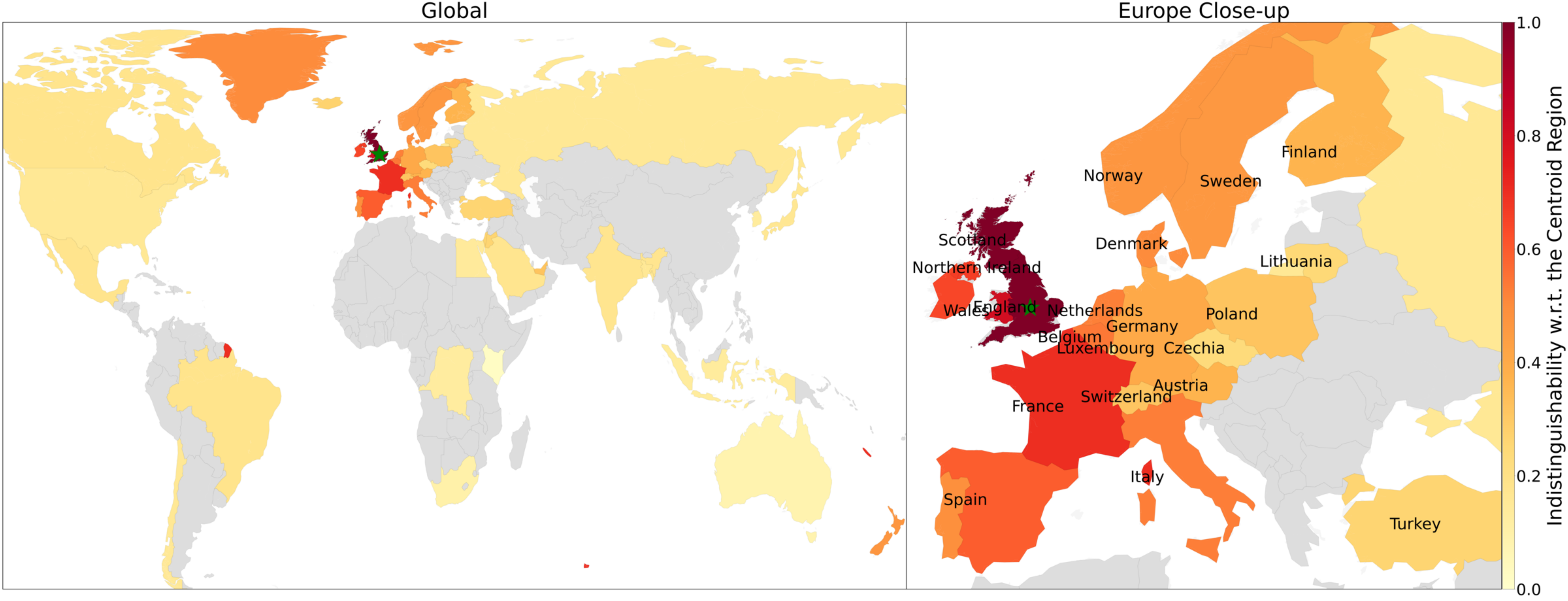}
        \caption{Centroid Region= England}
        \label{fig:plot3}
    \end{subfigure}
    \\
    \begin{subfigure}{1\textwidth}
        \centering
        \includegraphics[width=0.8\linewidth, height=1.85in]{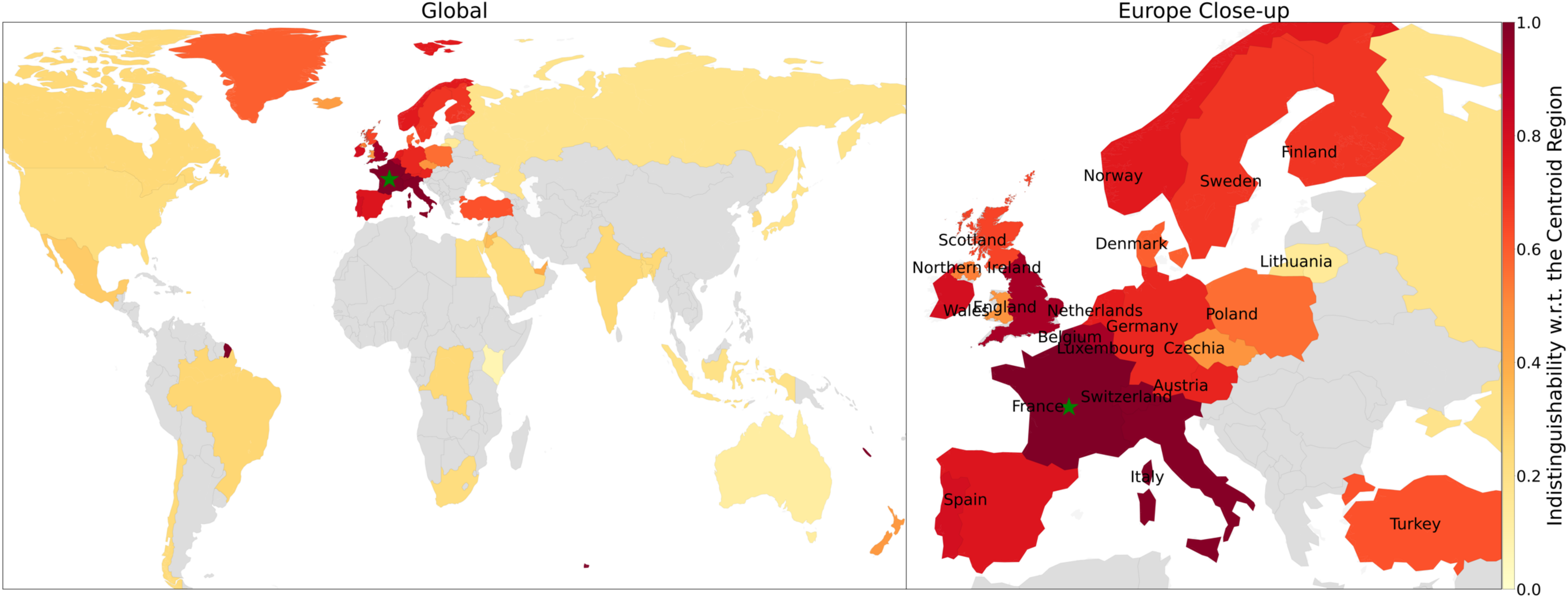}
        \caption{Centroid Region= France}
        \label{fig:plot4}
    \end{subfigure}
	\caption{Binary Naive Bayes Misclassification: The spatial dependencies among COVID genomes are also evident under the binary classification setting. Each subplot corresponds to a different centroid country, and a star misclassification network is constructed based on the Naive Bayes binary classification results relative to the centroid country. In (c) and (d), we are also showing the Europe close-ups of the misclassification networks on the right.}    
    \label{fig:Binary NB}
\end{figure}

We visualize the results using different centroid countries, with the redness of the regions indicating their indistinguishability to the centroid region. As shown in Fig.~\ref{fig:Binary NB}, for each selected centroid country, our results indicate that the majority of misclassifications occur between the centroid country and its neighboring countries. In North America, for example, genome sequences from the United States are most frequently misclassified as being from Canada and Mexico. In Asia, genomes from India are most frequently misclassified as those from Bangladesh and the UAE. In Europe, genome sequences from France are mostly misclassified as those from other neighboring European countries, such as Switzerland, Belgium, Luxembourg, and England. When France and England are used as the centroid countries, almost no misclassification occurs outside of Europe. For the UK, genomes from England are mostly misclassified with those from Northern Ireland, Wales, Scotland, Ireland, and France. Notably, Greenland is colored based on the results from Denmark, as it is considered Danish territory, and no experiments were conducted specifically using data from Greenland.

\subsection{True Misclassification vs. Configuration}
\begin{figure}[htbp]
\centering
\begin{subfigure}[b]{1\textwidth}
   \includegraphics[width=1\linewidth]{4c.pdf}
   \caption{NB Binary Classification}
    \label{fig:BinaryNBtrue}
\end{subfigure}

\begin{subfigure}[b]{1\textwidth}
   \includegraphics[width=1\linewidth]{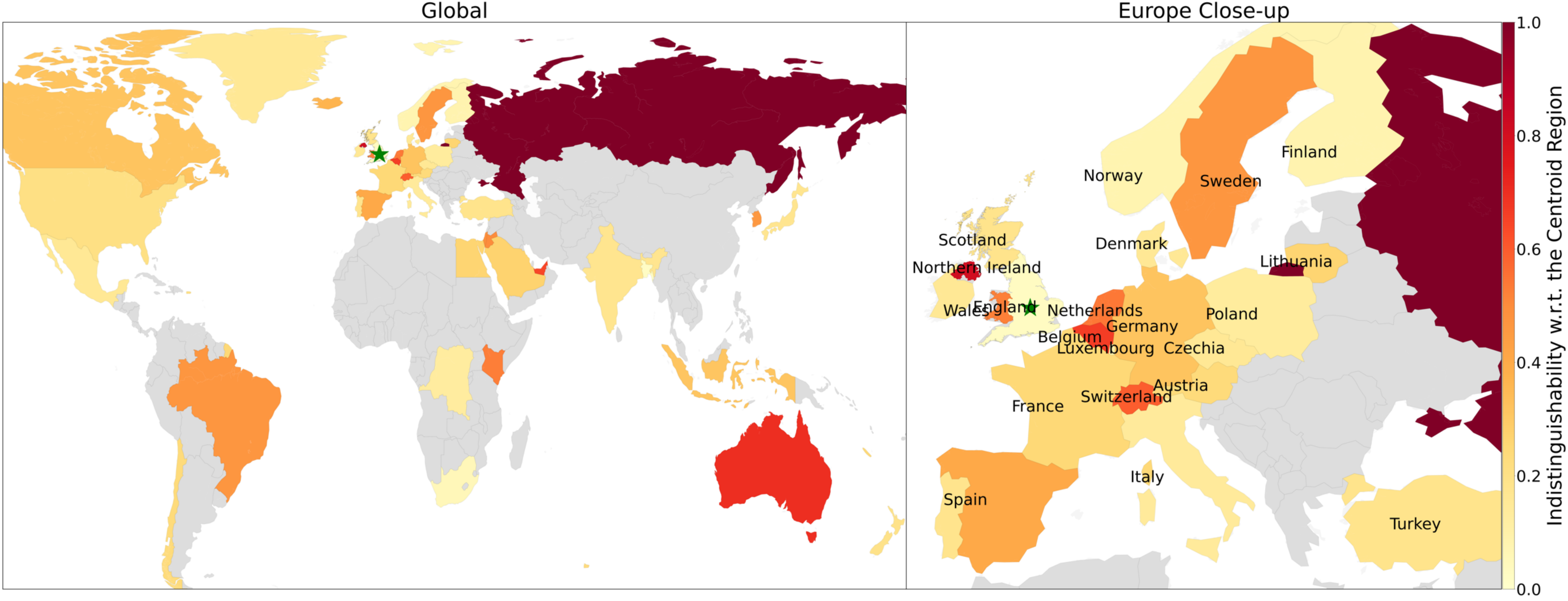}
   \caption{Configuration}
   \label{fig:BinaryNBconfig}
\end{subfigure}
\caption{Binary NB Misclassification Network vs. Configuration: The significance of the spatial dependencies is supported by the configuration model test. The misclassification network constructed under the binary scheme is a star network, relative to the centroid England. The redness of the other regions are determined by its indistinguishability score with resepct to genomes from England.}
\label{fig:BinaryNB}
\end{figure}

To further validate the significance of the spatial dependencies observed in the binary Naive Bayes (NB) misclassification network, we compared the true misclassification network with its corresponding configuration model in Fig.~\ref{fig:BinaryNBconfig}, similar to the comparison already shown for the multiclass setting in Fig.~\ref{subfig:MCCNBconfig}. Similarly, we observe that communities identified in the configuration graph show no spatial dependency. In Fig.~\ref{subfig:MCCNBconfig}, under the multiclass setting, the neighboring regions are often from different communities. And in the binary true misclassification network in Fig.~\ref{fig:BinaryNBtrue}, with England as the centroid, again, we observe that most of England genomes that got misclassified are distributed around its neighboring regions, most frequently, Scotland, Wale, Ireland and France. In general, the misclassified instances happened most frequently to the European regions and least frequently to North America and Asia. In the configuration graph, the most frequently misidentified regions are randomly scattered around the map as opposed to being close to the region of interest, England.

\subsection{Impact of International Travel on COVID Genome Variation}
\begin{figure}[htbp]
\centering
   \includegraphics[width=1\linewidth]{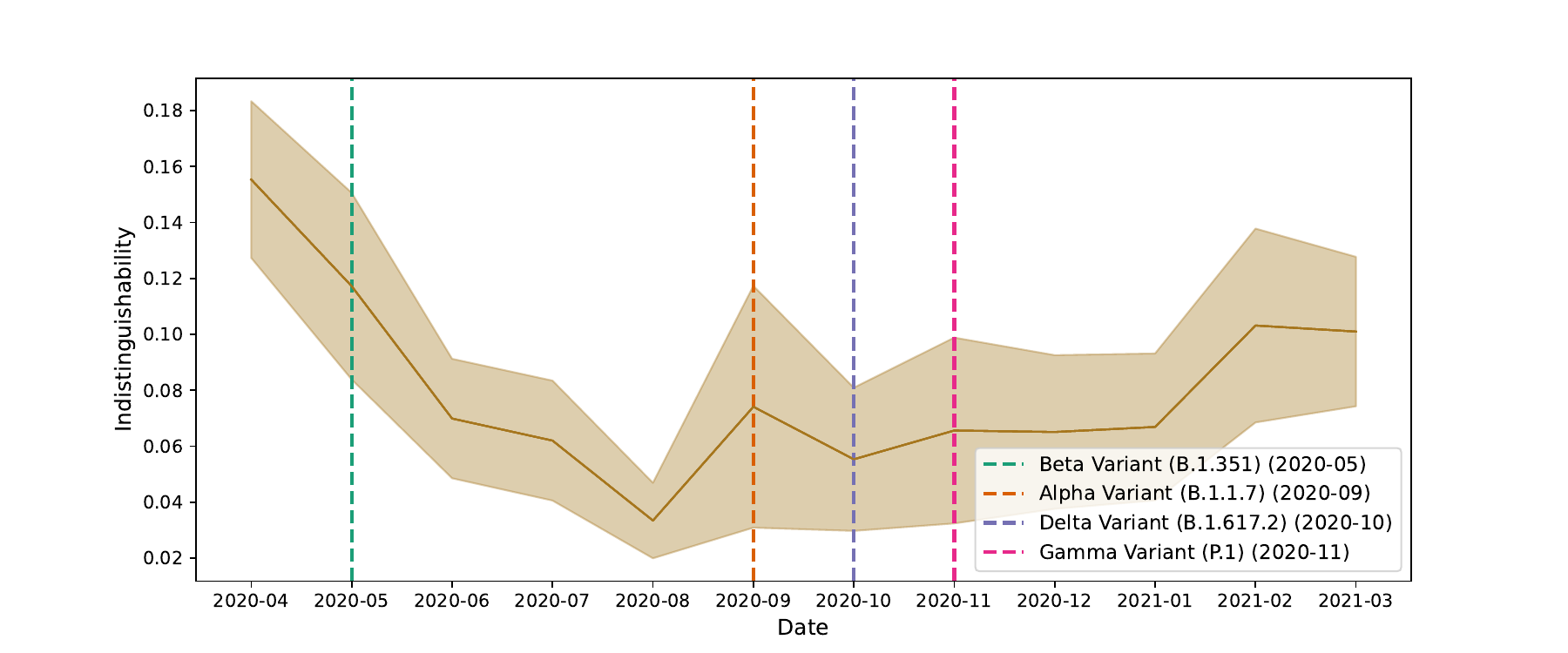}
\caption{Genome Indistinguishability with respect to Time and COVID VOC Occurrences: The figure shows the temporal relationship between genome indistinguishability and the emergence of Variants of Concern (VOCs). The indistinguishability of genome regions started at around 0.16 in April 2020 and declined steadily to 0.04 by August 2020, indicating the development of stronger regional genome characteristics. However, after the emergence of VOCs such as Alpha, Delta, and Gamma, the indistinguishability increased with fluctuation, suggesting increased genome variation, potentially due to the mixing of variants driven by international travel. The shaded area represents the 95$\%$ confidence interval of the accuracy score.}
\label{fig:TemporalAccuracy_and_COVID_VOC_occurrences}

\end{figure}

We investigated the impact of international travel activities on the evolution and variation of COVID genomes, by assessing the country-level aggregated indistinguishability of COVID sequences from both spatial and temporal perspectives. We quantified these impacts by examining the correlation between the centrality of countries in the OAG flight network and their respective country-level aggregated genome indistinguishability computed by misclassification rates. Our findings in Table ~\ref{tab:FlightsCentralitycorrelation} show significant correlations between travel activities and the aggregated pairwise genome sequence indistinguishability. The regional travel acitivities were assessed by their degree, eigenvector, closeness, and betweenness centralities within the flight network. The identified correlations suggest that countries with more central roles in international travel exhibit more complex COVID genome compositions, reflecting the influence of travel on genomic variations. Additionally, these results align well with the assumptions underlying the proposed GMNA framework, that travel and social contacts likely enhance homogeneity among epidemic genome ensembles from neighboring regions. This assumption could explain why genomes from more international regions exhibit higher aggregated indistinguishability. These results show the importance of taking the underlying transportation network into account when understanding and responding to the genomic evolution of COVID-19 and other epidemics. Further, using genomes from different time points, we examined the association between the emergence of Variants of Concern (VOCs) and the average aggregated genome indistinguishability scores, as depicted in Fig. \ref{fig:TemporalAccuracy_and_COVID_VOC_occurrences}.
Our results show that the accuracy of genome region prediction steadily increased at the beginning of COVID from around 0.84 in April 2020 to 0.96 in August 2020, suggesting developments of stronger regional characteristics during this period. The declined accuracy and hence increased aggregated genome indistinguishability after the emergence of Alpha, Delta, and Gamma variants could potentially be explained by variants mixing resulted by international travel.

\begin{table}[htbp]
\centering
\begin{tabular}{|l|c|c|}
\hline
\textbf{Centrality Measure} & \textbf{Spearman's Rank Correlation Coefficients} & \textbf{P-value} \\ \hline
Degree                      & 0.550                             & 2.55e-06         \\ \hline
Eigenvector                 & 0.584                             & 4.12e-07         \\ \hline
Closeness                   & 0.554                             & 2.05e-06         \\ \hline
Betweenness                 & 0.378                             & 2.05e-03           \\ \hline
\end{tabular}
\caption{Correlation between Centralities in the Flight Network and COVID Genome Sequence Indistinguishability: The identified correlations show that countries with more central roles in international travel have more complex COVID genome compositions, highlighting the effect of travel on genomic variations.}

\label{tab:FlightsCentralitycorrelation}
\end{table}

\begin{figure}[htbp]
\centering
   \includegraphics[width=0.9\linewidth]{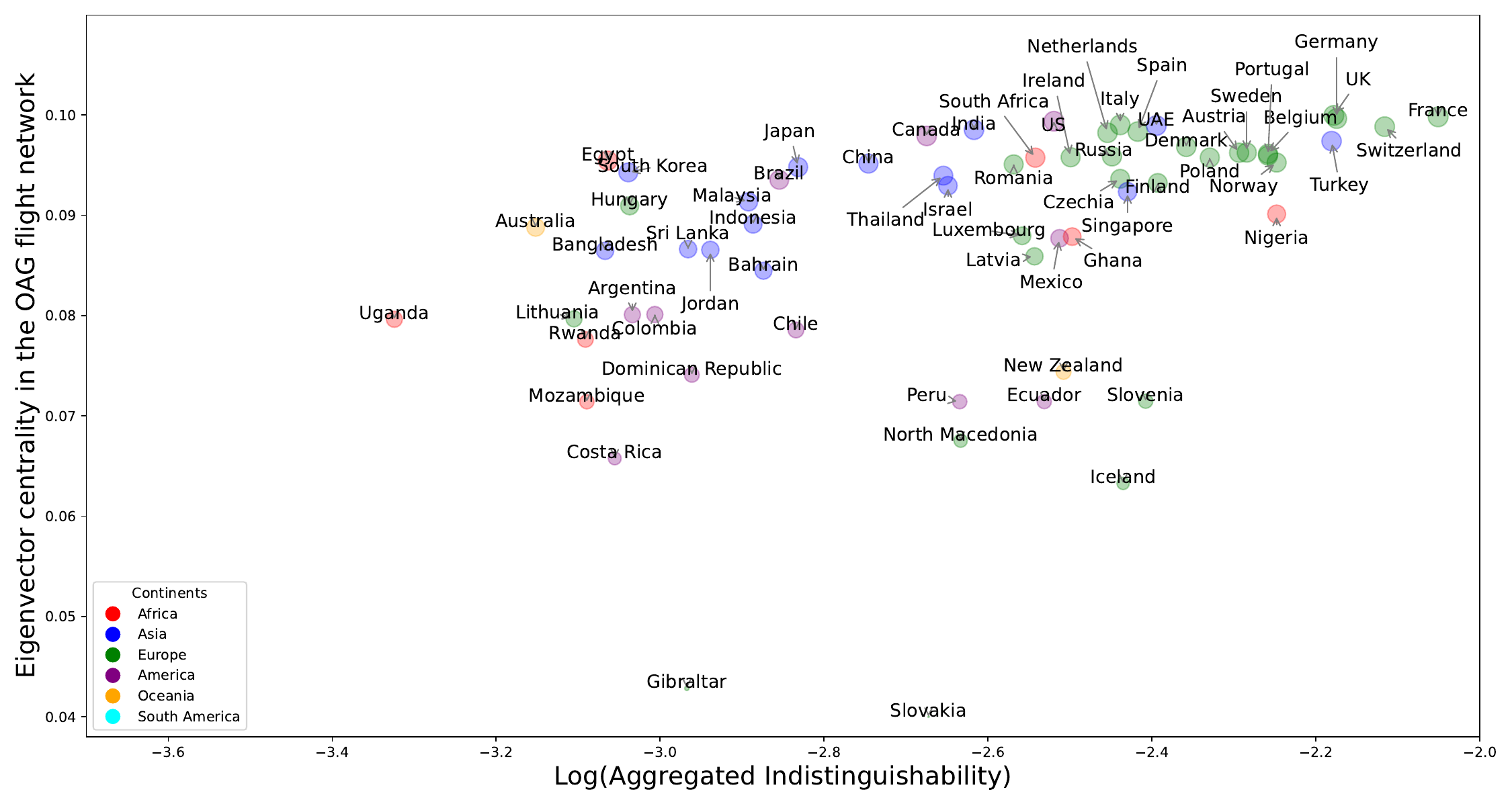}
\caption[]{Eigenvector Centrality vs. Log-Transformed Aggregated Indistinguishability of COVID Genome Sequences by Country: Countries are color-coded by continent. Countries with greater travel connectivity, particularly in Europe and Asia, tend to exhibit higher genome sequence indistinguishability.}
\label{fig:FlightsEigenvector_vs_indistinguishability_countrylevel}
\end{figure}

\subsection{Classification Accuracy}
In machine learning, improving performance of a trained classifier often serves as the primary research goal. While higher accuracy generally indicates a more reliable and credible model, it also results in limited data availability for misclassification-based inference. Consequently, a higher-accuracy classifier may not be the optimal option when the research objective is to learn from the misclassified instances.

For example, under the binary setting, a CNN classifier could easily achieve close to 100$\%$ classification accuracy. Although the misclassified instances produced by a higher-performance classifier intuitively suggest stronger class-wise associations, an overly accurate binary classifier that generates few misclassified instances offers limited insights into the genome associations. Moreover, the LOCO algorithm (see Section~\ref{sec:LOCO}) we designed to handle the trade-off between classification accuracy and the availability of misclassified data can only be applied to the multiclass classification setting, not the binary setting. Hence, for all experiments conducted under the binary setting, we use the Naive Bayes classifier to allow for more misclassified instances.

Similarly, under the multiclass classification setting, though from the perspective of model performance, deep learning-based models could classify genome sequences with higher accuracy, the results may be less informative for misclassification-based inference. This is because they require more data to train, resulting in datasets with reduced regional representation. Additionally, higher accuracy may result in outcomes with minimal gradient, with misclassified instances predominantly occurring in overt situations. As shown in Fig.~\ref{fig:CNN3000}, though the misclassification network from a more accurate CNN classifier still displays spatial dependency, the less inclusive dataset yields results that are comparatively less informative or comprehensive compared to Fig.~\ref{fig:BinaryNBtrue}. Hence, we emphasize that the main purpose of this project is not to push the classification accuracy of the employed classifier, but to show the applicability of using the misclassified instances for association inference.

We provide examples of model performance for reference in~Table \ref{tab:model_performance}. As described in \ref{subsec: Data and preprocessing},  we sample an equal number of genomes from countries whose data availability surpasses a certain threshold, resulting in a dataset with equal regional representation. The threshold value determines the number of regions for prediction. The sample size from each region and the size of the label set, representing the number of classes for the classification problem, directly impact the data size, computational complexity, and model performance.

\begin{table}[htbp]
  \begin{tabular}{cccccc}
    \toprule
    Model Name & Accuracy & Threshold & Sample Size & Label Set Size \\
    \midrule
    Binary CNN & 99.0\% & 1000 & 1000 & 33 \\
    Binary Naive Bayes (centroid=England) & 89.5\% & 100 & 100 & 33 \\
    Binary Naive Bayes (centroid=France) & 88.3\% & 100 & 100 & 33 \\
    Binary Naive Bayes (centroid=United States) & 92.5\% & 100 & 100 & 33 \\
    Binary Naive Bayes (centroid=India) & 95.6\% & 100 & 100 & 33 \\
    Multiclass CNN & 86.5\% & 10000 & 10000 & 9 \\
    Multiclass CNN & 75.6\% & 3000 & 3000 & 20 \\
    Multiclass CNN & 60.9\% & 1000 & 1000 & 33 \\
    Multiclass Naive Bayes & 64.9\% & 300 & 300 & 49 \\
    Enformer+CNN fine tune & 68.9\% & 10000 &10000  & 9 \\
    \bottomrule
  \end{tabular}
  \caption{Model Performance under Different Settings.  The table shows the accuracy of the models used in this study for reference. While higher accuracy is often a goal in machine learning, this study emphasizes that overly accurate classifiers may reduce the utility of misclassified instances for association inference. Classifiers with higher accuracy, such as CNNs, produce minimal misclassification, leading to association results with limited gradient and reduced detail. The results shown here highlight that the aim of our framework is not to maximize accuracy, but to use misclassification as a tool for exploring complex genomic relationships.}
\label{tab:model_performance}

\end{table}

\section{Discussion}
In this work, we present a novel alignment-free approach for comparative genomics that we term the genome misclassification network analysis (GMNA). Our framework is a generic network-generating method based on misclassification results for correlation-based network analysis. We introduce \textit{indistinguishability}, to quantify the association between pairs of genome groups and the genetic diversity of genome ensembles. We identify the pairwise association between the target outcome and the predicted outcome, which is then utilized to design a data-driven framework that can incorporate any state-of-the-art AI models for comparative genome analysis. We showed using more than 500,000 SARS-CoV-2 genomes that by employing GMNA, associations between sequences and geographic sampling location could be uncovered with limited computation resources.

Using our framework GMNA, we showed that SARS-CoV-2 genome sequences exhibit strong geographic clustering and the results are robust and consistent under various experimental settings. Further, the genome ensemble indistinguishabilities, which could indicate genetic variation and complexity, could be correlated to the centralities of their positions in a travel network, i.e. the OAG flight network. The centrality of a region in the flight network indicates its importance in global transportation. And this result suggests that human activities impact COVID genome variation. Additionally, we considered how the average indistinguishability of the genome sequences evolve during COVID and compared it with the emergence of the variants of concerns.

Previous efforts to classify SARS-CoV-2 genomes have focused on lineages tracking and identification using established comparative phylogenetic approaches \cite{forster2020phylogenetic,rambaut2020dynamic,tegally2021sixteen,zhou2021identification} and those based on k-mer clustering \cite{ali2021k, ali2021effective, washington2021emergence} have overall found high accuracy. Additionally, there have been some efforts leveraging neural networks to predict SARS-CoV-2 infection \cite{banerjee2020use,rosado2021multiplex,yang2020routine} or detect SARS-CoV-2 from different virus strains \cite{lalmuanawma2020applications,lopez2021classification,nachtigall2020detection}, which--in agreement with our findings--demonstrate the utility of such ML-based methodologies. These studies contribute to a growing body of literature on the utility of ML\slash AI models in studying emerging infectious disease outbreaks \cite{ardabili2020covid,islam2020systematic,schwalbe2020artificial,arora2021artificial,marcus2020artificial}. Our results extend from these approaches by highlighting the information contained in the misclassification network, which is relevant for both comparative and ML-based models. Future work should focus on linking comparative and ML approaches as a way of improving the explainability of the ML-models and investigating whether different ML approaches have favorable properties with respect to data training efficiency and\slash or generalizability.   

Our results demonstrating an association between travel and SARS-CoV-2 genome misclassification is not unexpected. Indeed, we see this results as validating our underlying modeling approaches. Previous phylogeographic studies of SARS-CoV-2 have found strong impacts of mobility on genome differentiation \cite{kraemer2021spatiotemporal,dudas2021emergence,lopez2021first,goliaei2024importations}. Coupled with studies of the effect of mobility on the underlying epidemiology of SARS-CoV-2 \cite{kraemer2020effect,chinazzi2020effect,wellenius2021impacts,rockett2020revealing,chang2021variation}, our findings contribute to a growing body of evidence showing the strong dependence of epidemic dynamics on mobility \cite{charu2017human,wesolowski2015impact,arenas2020modeling}. This finding, that mobility structures pathogen genomes and epidemic dynamics, has been found across a range of established \cite{charu2017human,lonnroth2017tuberculosis,peeters2015characterizing}, emerging \cite{zhang2017spread,alexander2015factors,kraemer2019utilizing,kraemer2020effect,chinazzi2020effect,wellenius2021impacts,rockett2020revealing,chang2021variation,peak2018population}, and re-surging \cite{wesolowski2015impact,kalipeni1998refugee} pathogens across a range of transmission routes and pathogen type, e.g., bacterial vs. viral. Again, that host mobility leaves a detectable signature on pathogen genomics has been well studied both theoretically \cite{sheppard2018population,boots2007local,castillo2016perspectives,poletto2015characterising} and empirically \cite{lemey2014unifying,poletto2013host,charu2017human} and is thought to be a quite general feature of population genetic differentiation \cite{petkova2016visualizing,loog2017estimating,leslie2015fine,wolf2017making,fischer2017estimating}.   

Regarding the choice of model for classification, our results show that Naive Bayes performed sufficiently well for genome comparative analysis. We include results incorporating a CNN-based model in the GMNA framework that demonstrate similar spatial dependency in the appendix~\ref{subfig:CNNhard}. Additionally, we employed a transfer learning approach by fine-tuning a pre-trained transformer-based \cite{vaswani2017attention} model, specifically the Enformer \cite{avsec2021effective}, for our genome classification task. However, contrary to our expectations, we did not observe an improvement in performance compared to the CNN model for this task. The classification performance from fine-tuning the Enformer is included in Table~\ref{tab:model_performance} for reference. We note that methods incorporating prior knowledge, such as negative sampling \cite{mikolov2013distributed,mikolov2013efficient}, have been shown to improve model performance by taking into account both similarity and dissimilarity in data context. However, we chose not to employ such techniques in our study, since the purpose of our study if to learn data associations, which would have been compromised if assumptions about data association were already incorporated during model training.

One additional limitation of this data association computation framework is that it only considers the association between groups of data and does not compare singular data points. However, we do not attempt to measure the association between two singular genomes or correlate such singular associations to their features, for the reason that the results would be lack of statistical significance. Indeed, the measured correlations and inferred associations between observable traits of individual sequences may not be reliably representative or comprehensive.
 
Overall, our study demonstrates the effectiveness of GMNA in identifying valuable information from misclassification data and providing insights into the spatial dependencies and potential factors contributing to the variation in SARS-CoV-2 genomes. Notably, misclassified instances during AI model training are often overlooked during results interpretation. Our proposed framework allows for the recycling of these misclassified instances, and our results show that misclassifications in genome sequencing data could provide insights into genetic variations associated with travel. We believe that repurposing misclassified instances holds promise in a wide range of applications beyond the scope of our study, such as healthcare, medical imaging, and diagnosis.

\section*{Code and Data Availability}
The code supporting the findings of this research is available in the GitHub repository, accessible at \url{https://github.com/wanhe13/Genome-Misclassification-Network-Analysis-GMNA-}. The repository includes all non-genomic data (see below for accessing those data), scripts, and instructions necessary to replicate the analyses presented in this study.

\begin{table}[htbp]
\centering
\begin{tabular}{@{}ll@{}}
\toprule
\textbf{GISAID Identifier:} & EPI\_SET\_20220727na \\
\textbf{DOI:} & \url{10.55876/gis8.220727na} \\
\bottomrule
\end{tabular}
\caption{\label{tab:sup-gisaid}\textbf{Data availability and Acknowledgements} - All genome sequences and associated metadata in this dataset are published in GISAID’s EpiCoV database. Anyone with valid GISAID Access Credentials can retrieve all records encompassed in the EPI\_SET ID. Those without GISAID Access Credentials may retrieve information about all contributors of data on which our analysis is based by either clicking on the DOI, or pasting the EPI\_SET ID in the "Data Acknowledgement Locator" on the GISAID homepage. To view the contributors of each individual sequence with details such as accession number, Virus name, Collection date, Originating Lab and Submitting Lab and the list of Authors, visit: \url{10.55876/gis8.220727na}.}
\end{table}

The reported Enformer+CNN model used the pretrained Enformer weights from \url{https ://huggingface.co/EleutherAI/enformer- official-rough}.

\section*{Acknowledgements} We would like to thank Professor Mortiz UG Kraemer for assistance acquiring and curating the genomic data and conversations about earlier iterations of this work.

\section{Supplementary Methods and Extended Results}
We included additional experimental results to further validate the robustness of the proposed Genome Misclassification Network Analysis (GMNA) framework. These experiments involved different AI accuracy levels and experimental setups, including soft and hard misclassification, as well as misclassification under the LOCO setup, to assess the consistency of the framework's results under diverse conditions.

\begin{figure}[htbp]
\centering
\begin{subfigure}[b]{1\textwidth}
   \includegraphics[width=1\linewidth]{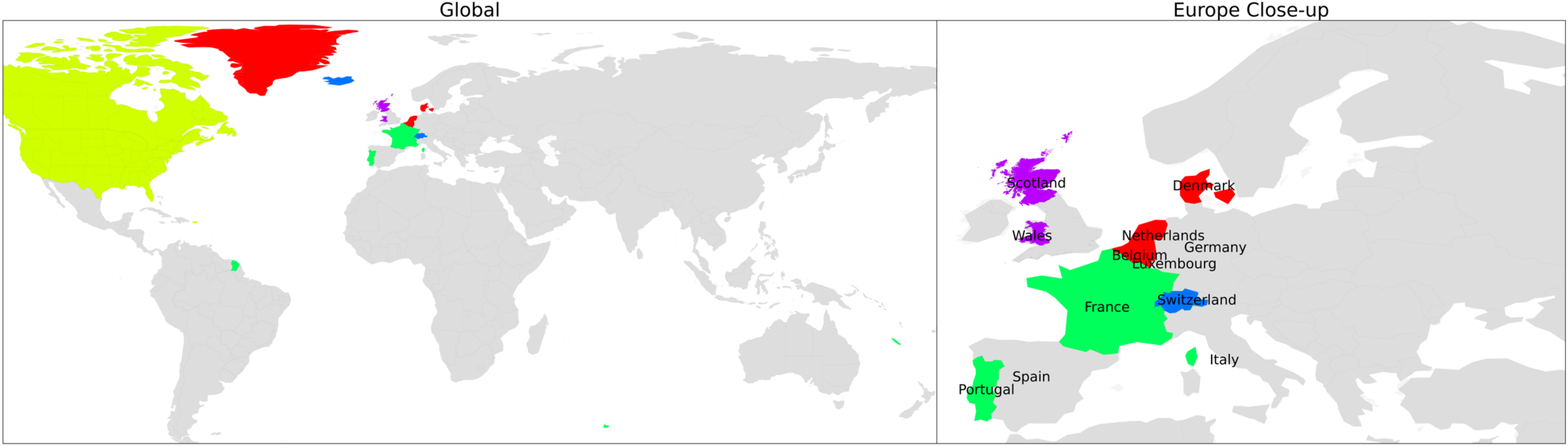}
\caption{LOCO Multiclass CNN Soft Misclassification Network: Clustering results on the soft misclassification network from a CNN LOCO classifier with a sampling threshold=sample size=3000.}
\label{subfig:LOCO}
\end{subfigure}

\begin{subfigure}[b]{1\textwidth}
   \includegraphics[width=1\linewidth]{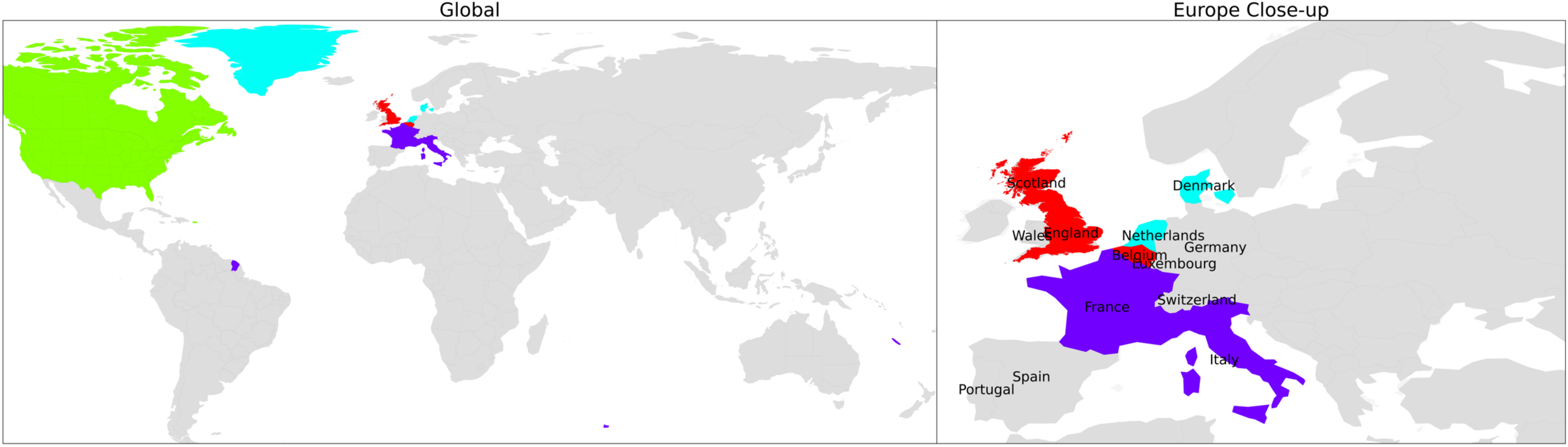}
   \caption{Multiclass CNN Soft Misclassification Network: Clustering results on the soft misclassification network from a CNN classifier with a sampling threshold=sample size=3000.}
\label{fig:CNN3000}
\end{subfigure}
\caption{Multiclass CNN Soft Misclassification Network: Clustering results on the soft misclassification network from a convolutional neural network classifier with a sampling threshold=sample size=3000.}
\label{fig:3000_LOCO_soft}

\end{figure}

In Fig.~\ref{fig:3000_LOCO_soft}, we observed spatial dependencies among COVID-19 genomes from neighboring regions, consistent with our previous findings. However, in this setting, with the higher prediction accuracy of 74.12\% under MCC, and 72.78\% under the LOCO setup, the spatial dependencies were confined to Europe and North America, where sufficient misclassifications were available for correlation analysis.

\begin{figure}[htbp]
\centering
\begin{subfigure}[b]{1\textwidth}
   \includegraphics[width=1\linewidth]{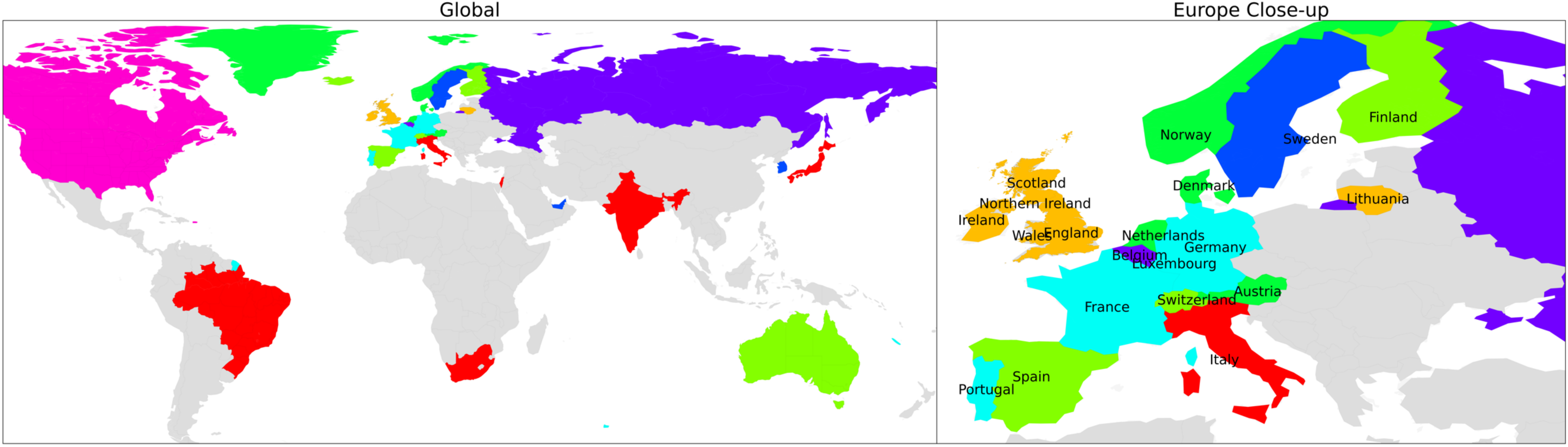}
   \caption{CNN Hard Classification}
   \label{subfig:CNNhard}
\end{subfigure}

\begin{subfigure}[b]{1\textwidth}
   \includegraphics[width=1\linewidth]{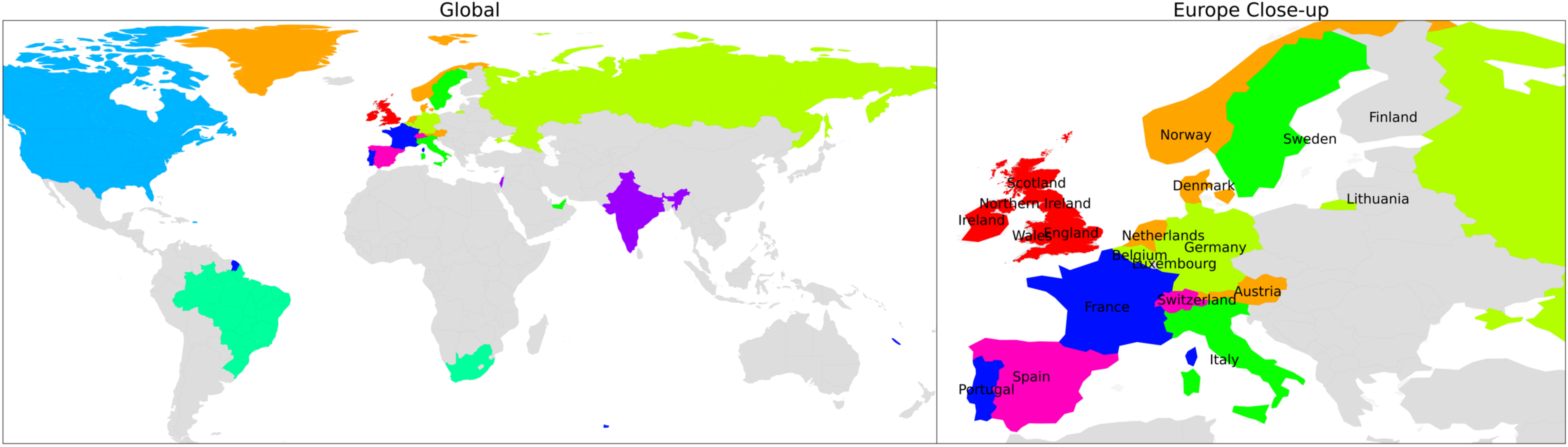}
   \caption{CNN Soft Classification using the Entire Output Distribution from SoftMax}
\end{subfigure}
\caption{Multiclass CNN Misclassification Network: Clustering results on the misclassification network from a convolutional neural network classifier with a sampling threshold=sample size=1000.}
\label{fig:hard vs. soft}
\end{figure}

Unlike in Fig.~\ref{fig:MCCNB_network} where the classification accuracy is lower at 58.02\% and in Fig.~\ref{fig:hard vs. soft} at 60.88\%, more misclassified instances were resulted from the less accurate classifier and gave us more instances for correlational analysis on genome ensemble association at a global scope. 

These supplementary results confirm that while higher prediction accuracy reduces the number of misclassified instances available for correlation analysis, the GMNA framework remains effective in detecting region-specific dependencies when sufficient data are available. This demonstrates the flexibility of the framework to adapt to AI's with varying accuracy while still providing meaningful insights into the spatial associations of COVID-19 genome sequences.

\end{document}